\documentclass[twocolumn]{aastex701}

\usepackage{graphicx}	
\usepackage{amsmath}	
\usepackage{amssymb}	
\usepackage{threeparttable}
\usepackage{comment}
\usepackage{caption}

\def\XMM        {{\em XMM}\/}
\def\XMM        {{\em XMM-Newton}\/}

\begin{document}

\title{A merger shock traced by radio arcs and ultra-long radio tails in galaxy cluster A2142}


\author[orcid=0000-0003-0628-5118]{Chong Ge}
\affiliation{Department of Astronomy, Xiamen University, Xiamen, Fujian 361005, China}
\email[show]{chongge@xmu.edu.cn}  

\author[orcid=0000-0001-5880-0703]{Ming Sun} 
\affiliation{Department of Physics \& Astronomy, University of Alabama in Huntsville, 301 Sparkman Dr NW, Huntsville, AL 35899, USA}
\email{fakeemail1@google.com}

\author[orcid=0000-0002-0775-6017]{Chris Nolting}
\affiliation{Gustavus Adolphus College, Saint Peter, Minnesota, USA}
\email{fakeemail4@google.com}

\author[orcid=0000-0002-9112-0184]{Fabio Gastaldello}
\affiliation{INAF–IASFMilano, via A. Corti 12, 20133 Milano, Italy}
\email{fakeemail2@google.com}

\author[orcid=0000-0001-7917-3892]{Dominique Eckert}
\affiliation{Department of Astronomy, University of Geneva, Versoix CH-1290, Switzerland}
\email{fakeemail5@google.com}

\begin{abstract}
Abell 2142 (A2142) is a massive, nearby galaxy cluster undergoing a complex merger.
It exhibits an elongated X-ray morphology along the northwest-southeast axis and hosts four known cold fronts.
Using XMM-Newton observations, we detect a merger shock on the northwest side of the cluster with a Mach number of $\mathcal{M} \sim 1.3$. 
The observed shock front and four cold fronts can be reproduced by numerical simulations of an off-axis merger with a large impact parameter, which imparts significant angular momentum to induce the sloshing of the subcluster core and large-scale ambient gas.
In projection, the shock front is spatially coincident with arc-shaped radio filaments observed behind the prominent head-tail radio galaxies T1 and T2. We interpret these radio arcs as partial vortex ring structures (resembling ``smoke rings'') produced by the interaction of the merger shock with the low-density cocoons of radio galaxies. The shock strips and rolls the jet cocoon into a toroidal vortex, as predicted by recent magnetohydrodynamic simulations. We further demonstrate that the merger shock can significantly elongate the radio tails by re-accelerating aged relativistic electrons and stretching the tail plasma via the post-shock wind. This process provides a natural explanation for the $>$500~kpc tail observed in this and other merging clusters. Our findings establish radio arcs and ultra-long radio tails as independent, complementary tracers of merger shocks in galaxy clusters. Our results demonstrate that merger shocks can reshape both the thermal and non-thermal components of galaxy clusters, and that tailed radio galaxies serve as sensitive probes of intracluster medium weather.

\end{abstract}


\section{Introduction}

Galaxy clusters form hierarchically through the accretion and merger of smaller structures at the nodes of the cosmic web. During cluster mergers, enormous amounts of gravitational potential energy are dissipated into the intracluster medium (ICM), driving shocks, turbulence, and particle acceleration \citep[e.g.,][]{markevitch2007}. These merger shocks compress and heat the ICM, elevate its pressure, significantly influence the evolution of cluster galaxies. The shocks may trigger and enhance star formation, active galactic nucleus (AGN) activity, and ram pressure stripping \citep[e.g.,][]{Owers2012,Roediger2014,Sobral2015}.

Merger shocks are typically detected through discontinuities in X-ray surface brightness and temperature profiles \citep[e.g.,][]{markevitch2007}. However, they also leave distinct imprints on the non-thermal components of clusters. On one hand, shocks can directly accelerate or re-accelerate relativistic electrons via diffusive shock acceleration (DSA), giving rise to large-scale, elongated synchrotron sources known as radio relics. These relics are often found in cluster peripheries and serve as unambiguous tracers of merger shocks \citep[e.g.,][]{vanWeeren2019}. On the other hand, shocks can dramatically reshape the morphologies of pre-existing radio galaxies. When a shock encounters the low-density cocoon of a radio galaxy, the shock propagates rapidly through the cavity, generating strong shear along the cavity boundary and transforming the cocoon into a toroidal vortex ring, like a ``smoke ring" \citep[e.g.,][]{Enblin2002,Pfrommer2011,Ge2026}. If the AGN jets remain active, the post-shock wind can deflect or even reverse the jets, producing asymmetric narrow-angle tailed (NAT) or head-tail morphologies \citep[e.g.,][]{Jones2017, Nolting2019}. Thus, radio galaxies can act as ``weather vanes" for ICM dynamics, providing unique constraints on shock properties and flow patterns.

\begin{figure*}
    \centering  
\includegraphics[angle=0,width=0.99\textwidth]{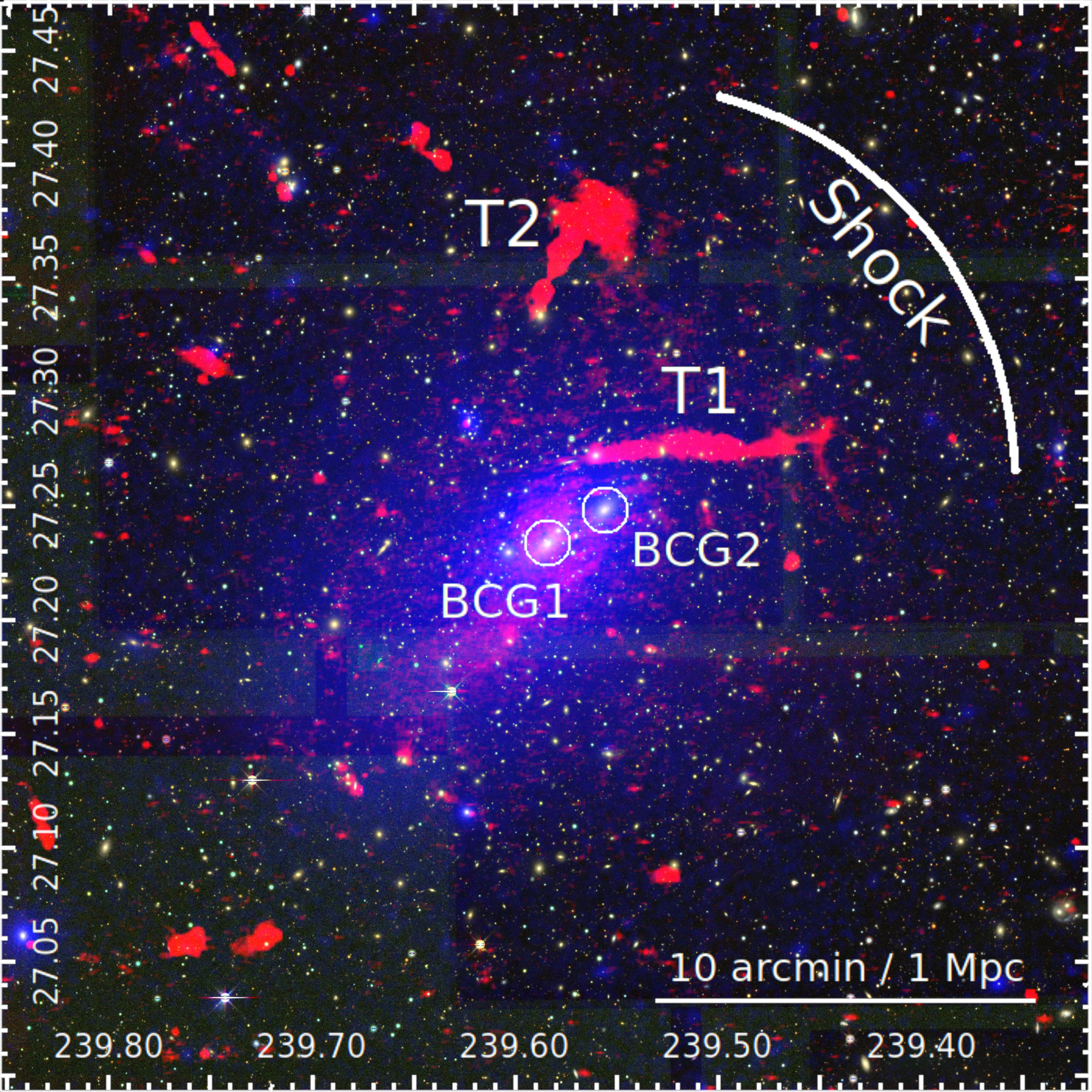}    
\caption{Optical image from the Dark Energy Camera Legacy Survey (DECaLS; \citealt{Dey2019}) in the $g, r, z$ bands, overlaid with a red radio image from LoTSS-DR3 144 MHz \citep{Shimwell2026}, and a blue X-ray image from XMM-Newton 0.5-2 keV mosaic. The white circles mark two BCGs. The white arc depicts the shock front detected from the X-ray data. Two head-tail radio galaxies T1 and T2 are located near the shock front, which may stretch the radio tails and meanwhile induce the arc-shaped radio filaments behind the ends of these radio tails.}
\label{fig:rgb}  
\end{figure*} 

Abell 2142 (A2142; Figure~\ref{fig:rgb}) is a nearby ($z=0.08982$; \citealt{Liu2018}) massive cluster ($M_{200} \approx 1.3\times 10^{15}\,M_\odot$; \citealt{Munari2014}) located at the center of the A2142 supercluster \citep[e.g.,][]{Einasto2015}. It is a merging galaxy cluster based on multiwavelength observations. The two brightest cluster galaxies (BCGs) show a large line-of-sight velocity difference of $\sim 1600$ km s$^{-1}$, a strong indicator of an ongoing merger, as they are the dominant members of the merging subclusters. Moreover, the BCGs align with the ICM elongation along the northwest–southeast (NW-SE) axis, which is also coincident with the orientation of the larger-scale supercluster filament \citep[e.g.,][]{Henry1996,Einasto2015}. Chandra observations revealed three prominent cold fronts in its core \citep{Markevitch2000,Wang2018}, making A2142 the first object in which cold fronts were discovered by Chandra. Subsequent XMM-Newton observations uncovered a fourth cold front at $\sim 1$ Mpc to the SE \citep{Rossetti2013}. 
The X-ray temperature maps and profile show a higher temperature in the NW side \citep{Henry1996,Owers2009,Rossetti2013,Tchernin2016}, as evidence of shock heating. 
Meanwhile, the particularly high Sunyaev-Zel'dovich (SZ) signal in the NW also suggests shock compression of the ICM \citep{Umetsu2009}. 
Optical spectroscopy reveals numerous substructures and ongoing mergers \citep[e.g.,][]{Owers2011,Liu2018}, while weak lensing maps indicate a disturbed mass distribution similar to the Bullet cluster \citep[e.g.,][]{Okabe2008}.
In the radio band, A2142 hosts a spectacular multi-component radio halo \citep[e.g.,][]{Venturi2017,Bruno2023,Riseley2024} extended along the NW-SE axis as an indication of cluster merger activities, and two prominent head-tail radio galaxies with ultra-long tails, T1 ($>$500~kpc) and T2 ($>$400~kpc), also aligned close to this axis. Their complex morphologies indicate strong interactions with the ICM \citep[e.g.,][]{Bruno2024}. Specifically, LOFAR observations reveal arc-shaped radio filaments extending beyond the tails of T1 and T2 \citep[e.g.,][]{Riseley2024,Bruno2024}, suggestive of shock-induced vortex structures.

In this paper, we report the discovery of a merger shock on the NW side of A2142, supported by both X-ray surface brightness and temperature discontinuities. We show that this shock is spatially coincident with the radio arcs behind T1 and T2, and we interpret these arcs as partial vortex rings produced by shock-cocoon interactions. We further demonstrate that the shock plays a critical role in elongating the radio tails, and argue that ultra-long radio tails can be used as independent tracers of merger shocks in galaxy clusters. We adopt a $\Lambda$CDM cosmology with $H_{0} = 70~\mathrm{km~s^{-1}~Mpc^{-1}}$, $\Omega_{m} = 0.3$, and $\Omega_{\Lambda} = 0.7$. At the cluster redshift of $z = 0.08982$, $1^{\prime \prime} = 1.676$~kpc.

\section{Data analysis}
\label{sec:data}
\begin{figure*}
\begin{center}
\centering
\includegraphics[angle=0,width=0.497\textwidth]{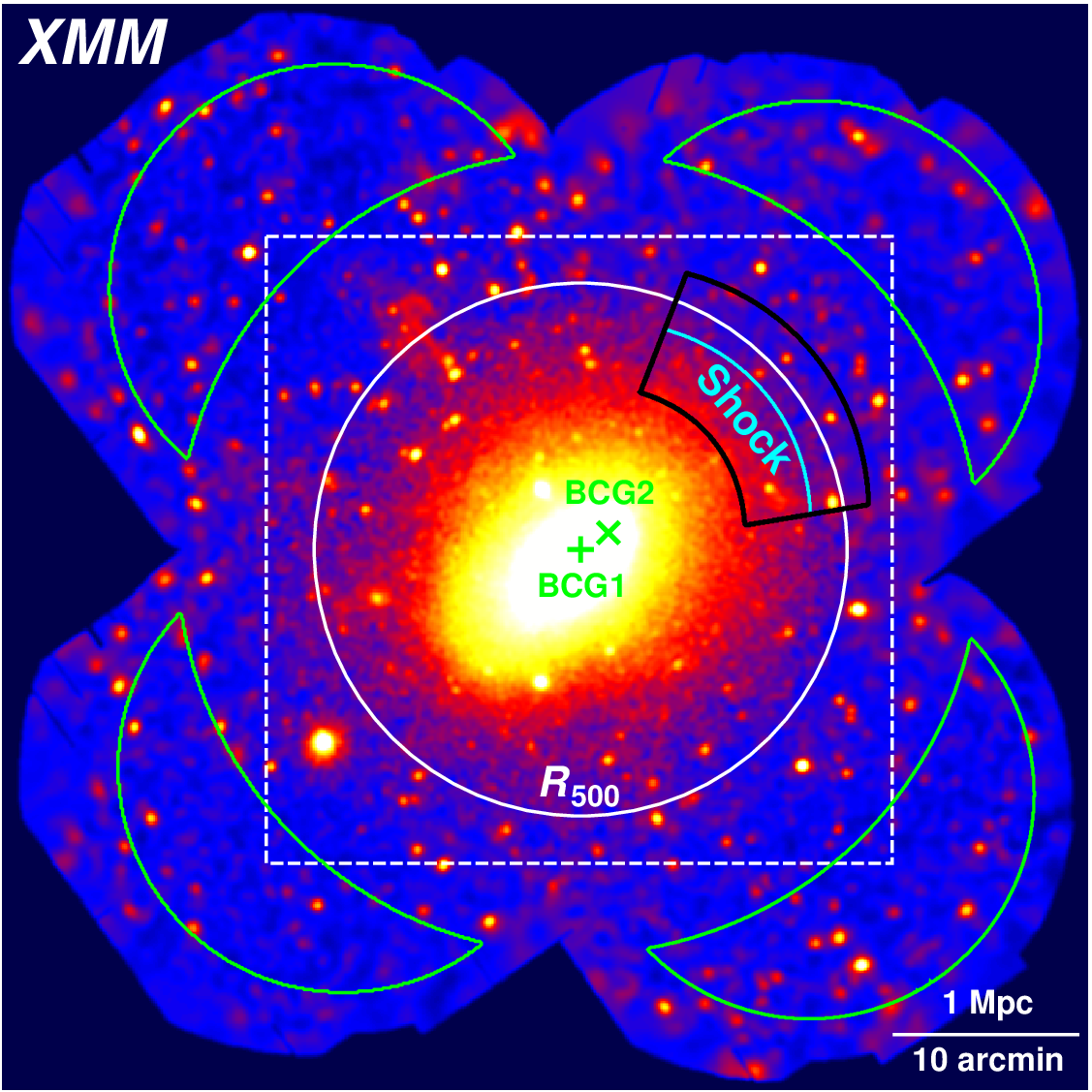}
\includegraphics[angle=0,width=0.497\textwidth]{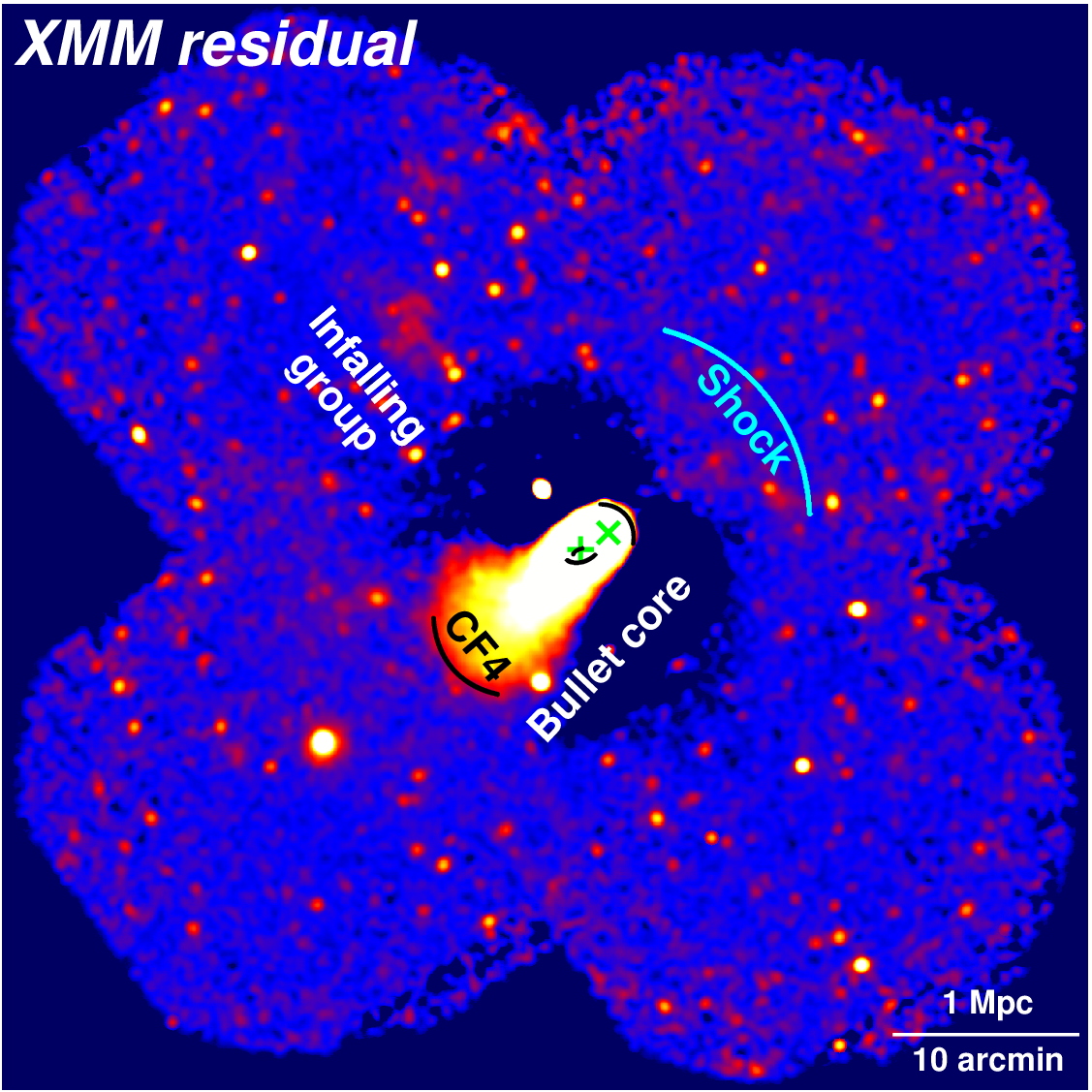}
\includegraphics[angle=0,width=0.497\textwidth]{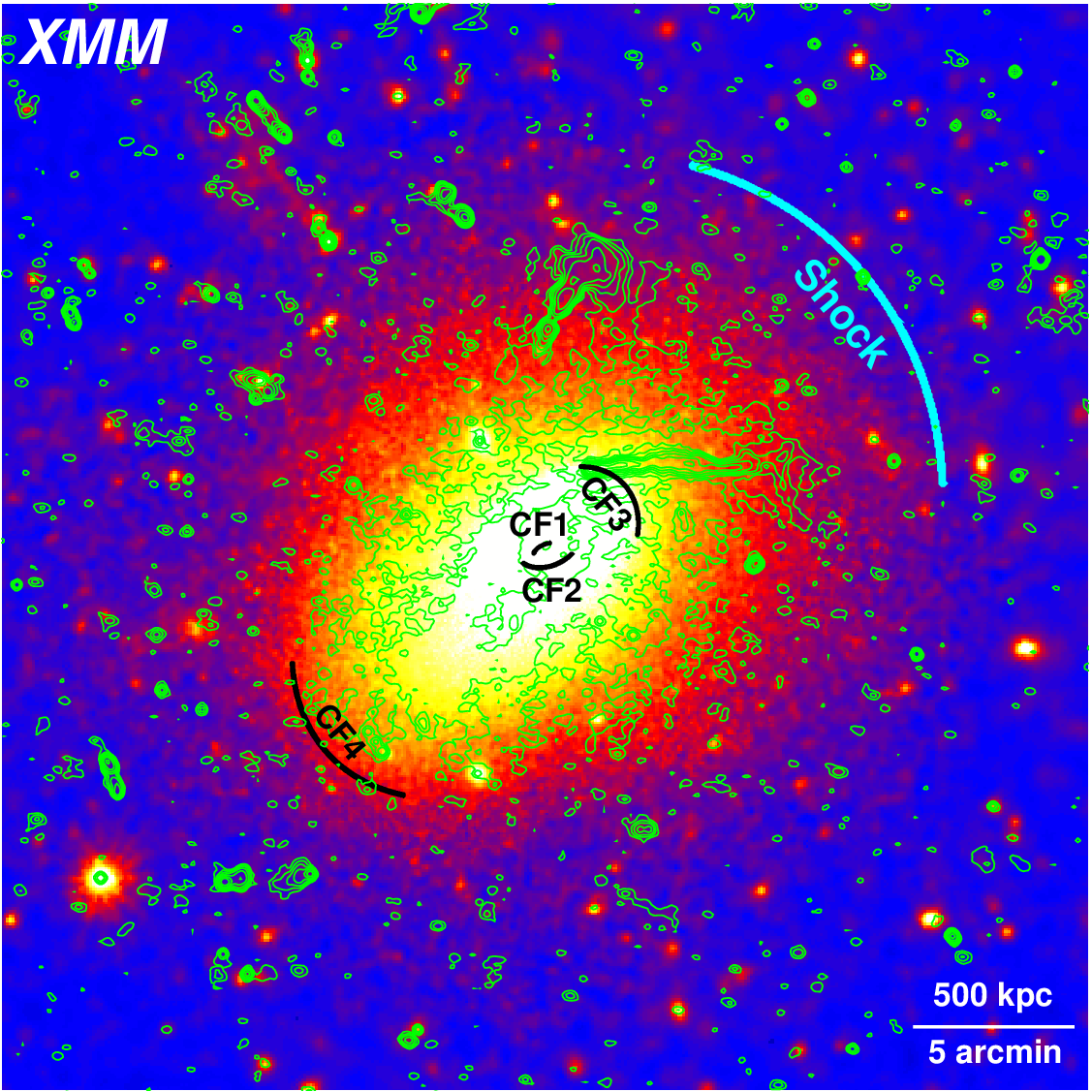}
\includegraphics[angle=0,width=0.497\textwidth]{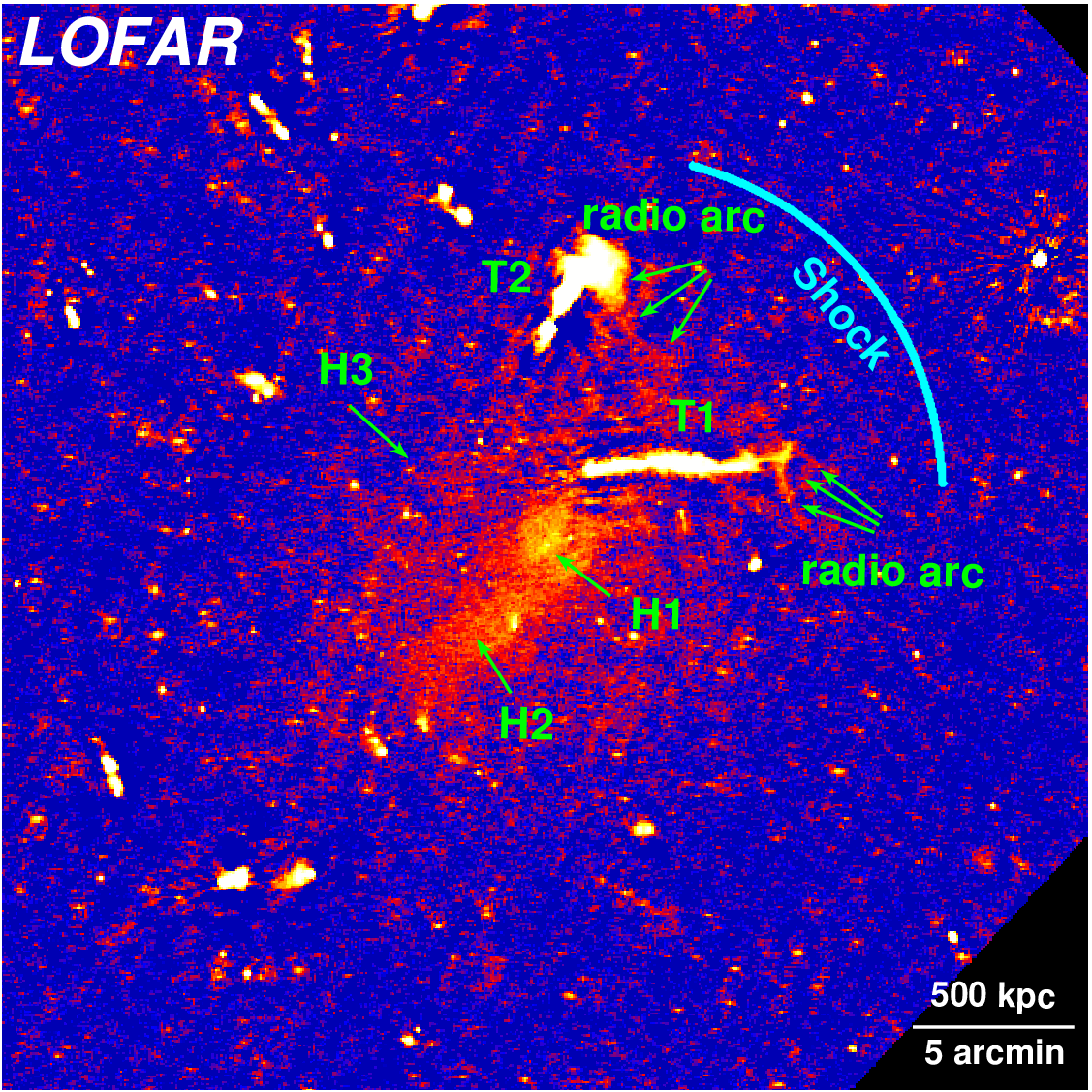}

\vspace{-0.1cm}
\caption{
{\bf Top left}: XMM-Newton 0.5-2 keV mosaic of A2142. The white circle marks $R_{500} = 14.2'$, and the white dashed box shows the center region enlarged in the bottom panels. The green sectors are the background regions. The black sector marks the extraction region for the SBP and the temperature profile shown in Figure~\ref{fig:rp}. The cyan line marks the location of the shock front. The green plus and cross mark the BCG1 and BCG2, respectively. 
{\bf Top right}: XMM-Newton residual image with the best-fit ICM model subtracted. Prominent residual features include the central bullet core and the infalling group of galaxies in the NE. The locations of four known cold fronts (CFs) are marked with black arcs.
{\bf Bottom left}: XMM-Newton 0.5-2 keV image of A2142 within $R_{500}$. The green contours are from LoTSS-DR3 144 MHz image \citep{Shimwell2026} showing the multi-component radio halo and tailed radio galaxies. Black arcs mark four CFs. We find a merger shock front (marked by the cyan arc) traced by radio arcs and ultra-long tails of tailed radio galaxies. 
{\bf Bottom right}: LoTSS-DR3 144 MHz \citep{Shimwell2026} image shows the multi-component radio halo (H1, H2, and H3; \citealt{Bruno2023}) and tailed radio galaxies (T1 and T2; \citealt{Bruno2024}). The radio arcs are likely partial vortex ring structures produced by a cluster merger shock, because the shock can strip and roll the jet cocoon into a vortex ring structure, resembling a “smoke ring” behind the end of the jet.
}
\label{fig:img}
\end{center}
\end{figure*}

We analyze archival XMM-Newton observations of A2142 (Table~\ref{t:obs}), including a central pointing (ObsID = 0674560201; \citealt{Rossetti2013}) and four offset pointings from the XMM Cluster Outskirts Project (X-COP; \citealt{Eckert2014,Tchernin2016}) with an additional pointing (ObsID = 0111870301) covering the shock region at the NW side. The data are reduced using the Extended Source Analysis Software (ESAS) integrated into the XMM-Newton Science Analysis System (SAS v17.0.0), following the procedures described in \cite{Ge2019a}. 
The reduced background-subtracted, exposure-corrected, and smoothed mosaics of A2142 are shown in Figure~\ref{fig:img}.

For spectral analysis, we utilize the double-background subtraction method to subtract the instrumental non-X-ray background and X-ray backgrounds separately \citep{Ge2021b}.
The instrumental non-X-ray background (NXB) spectra are extracted from the XMM-Newton filter-wheel closed data sets. These rescaled NXB spectra are then loaded into XSPEC as the background spectra. 
The X-ray sky components are extracted from the off-source regions (green sectors in Figure~\ref{fig:img}) and properly modeled in XSPEC. 
The X-ray sky backgrounds are represented by an unabsorbed APEC (from the local hot bubble), an absorbed APEC (Galactic background), and an absorbed POWERLAW (AGN-dominated background). 
We further constrain these components by jointly fitting them with an off-cluster ROSAT All Sky Survey (RASS) spectrum extracted from a 0.5 deg to 1.5 deg annulus surrounding the cluster.
The sky contributions of on-source and off-source regions are then linked by their sky solid angles. The cluster X-ray emission in the on-source region is fitted with an additional absorbed APEC model.
We use the solar abundance table of \cite{Asplund09} in the spectral fits, and adopt an absorption column density of $N_{\rm H} = 4.36 \times 10^{20}\ {\rm cm}^{-2}$ from NHtot tool \citep{Willingale13}.

\begin{table}
\centering
\setlength{\tabcolsep}{1.5pt}
\caption{\XMM\ Observations}
\begin{tabular}{lcccc}
\hline
ObsID & PI & Date Obs & Exp$^a$ (ks) & Clean exp (ks)\\
\hline
0111870301 & F. Jansen & 2002-09-10 & 20.3/17.9 & 7.1/3.9\\
0674560201 & M. Rossetti &2011-07-13 & 59.1/53.4&53.2/41.1\\
0694440101 & D. Eckert & 2012-07-14 & 24.5/21.0&18.3/10.0\\
0694440201 & D. Eckert  & 2012-07-14 & 34.6/30.7&32.8/25.3\\
0694440501 & D. Eckert  & 2012-07-16  & 34.6/30.7&32.1/25.9\\
0694440601 & D. Eckert  & 2012-07-18 & 38.6/34.7&31.2/20.2  \\
\hline
\end{tabular}
\begin{tablenotes}
\item
$^a$: Exposure time for MOS/pn.
\end{tablenotes}
\label{t:obs}
\end{table}

\section{results}
\label{sec:result}

\subsection{ICM properties}
\label{sec:icm}

A2142 is the richest maxBCG cluster \citep{Koester2007} and is included in the sample of \cite{Ge2019a} to study X-ray scaling relations of optical selected clusters. The spectroscopic X-ray temperature $T_X = 8.20\pm0.05$ keV is measured in $0.15-0.75 R_{500}$ \citep{Ge2019a} , with $R_{500}$ being estimated from the $M-T_X$ relation \citep{Sun2009} iteratively. The derived cluster radius and mass are $R_{500}/R_{200} = 14.2/21.2$ arcmin (1.4/2.1 Mpc) and $M_{500} / M_{200} = 9.1\times 10^{14}/1.2 \times 10^{15}\ M_\odot$, which is consistent with the optical and SZ measurements \citep[e.g.,][]{Munari2014,Planck2016}.
Its global ICM distribution can be modeled with a $\beta$-model \citep{1976A&A....49..137C}.
We extract surface brightness profiles (SBPs) excluding regions of point sources and diffuse substructures, centered on BCG1 (Figure~\ref{fig:rgb}; R.A. = 239.58, decl. = 27.23), the center of the large-scale X-ray emission and gravitational potential \citep{Wang2018}. 
The $\beta$-model gas distribution is given by
\begin{equation}
n_{\rm ICM}(r) =n_{\rm e0}\left(1+\frac{r^2}{r_c^2}\right)^{-\frac{3}{2}\beta},
\end{equation} 
with the derived X-ray SBP also following a $\beta$-model in the form of 
\begin{equation}   
I_{\rm X}(r) =I_0\left(1+\frac{r^2}{r_c^2}\right)^{\frac{1}{2}-3\beta}.
\end{equation}  
We apply the analytical formula of \cite{Ge2016} to convert the central surface brightness $I_0$ (from the $\beta$-model fitting of the SBP) to the central electron density $n_{\rm e0}$.
The resulting best-fit parameters are listed below:
\begin{equation*}  
\begin{aligned}  
& n_{\rm e0}=(3.24\pm0.05)\times 10^{-3}{\rm\ cm^{-3}}, \\
& I_0=(2.10\pm0.04)\times 10^{-5}{\rm\ counts\ s^{-1}\ arcsec^{-2}}, \\ 
& r_c=260.7\pm5.5 {\rm\ arcsec},\ \beta=0.84\pm0.01.
\end{aligned}  
\end{equation*} 

To reveal the substructures, we subtract the aforementioned best-fit $\beta$-model from the XMM-Newton mosaic to produce the residual images in Figure~\ref{fig:img}, following procedures in \cite{Ge2021b}. Notable substructures include the infalling group of galaxies in the NE \citep[e.g.,][]{Eckert2014}, and the bullet core, which is also revealed in the residual XMM-Newton image by \cite{Rossetti2013} using a different method. 
The smooth outskirts residual, aside from the bullet core, suggests that A2142 is undergoing an intermediate-mass-ratio merger event between the major and minor merger \citep[e.g.,][]{Rossetti2013,Wang2018,Riseley2024}.
This intermediate-mass-ratio merger scenario is further supported by numerical simulations presented in Section~\ref{sec:scenario}. Although less disruptive than a major merger, this event can still leave a merger remnant in the cluster core and drive a shock front into the outskirts.

\subsection{Merger shock}
\label{sec:shock}
The merger model of A2142 proposed in \cite{Markevitch2000} suggests a merger shock ahead of the bullet core.  Meanwhile, the higher X-ray temperature and SZ signal in the NW compared with other directions indicate shock heating and compression \citep[e.g.,][]{Owers2009,Umetsu2009,Tchernin2016}.
Moreover, the radio arcs behind T1 and T2 resemble those observed behind the tailed radio galaxy ESO~137-007 in A3627 \citep{Koribalski2024,Ge2026}. These radio arcs are likely partial vortex ring structures that originate from the jet cocoons, which are stripped and rolled by a shock \citep[e.g.,][]{Enblin2002,Jones2017,Nolting2019}. 
The observed arcs exhibit more irregular and wispy morphologies compared to idealized simulated vortex rings, likely because numerical simulations often neglect the effects of turbulence and buoyancy, which can distort and disrupt the coherent ring structure. In this respect, the wispy, uneven surface brightness ring associated with the tailed radio galaxy NGC 1265 \citep[e.g.,][]{Pfrommer2011} provides a striking analog, further supporting the interpretation that these arcs are indeed partial vortex rings shaped by the shocks and the ambient turbulent ICM.
To search for this potential shock, we extract a SBP in a sector located near the radio arcs behind T1 and T2, as shown by the black sector in Figure~\ref{fig:img}.

\begin{figure}
\begin{center}
\centering
\includegraphics[angle=0,width=0.45\textwidth]{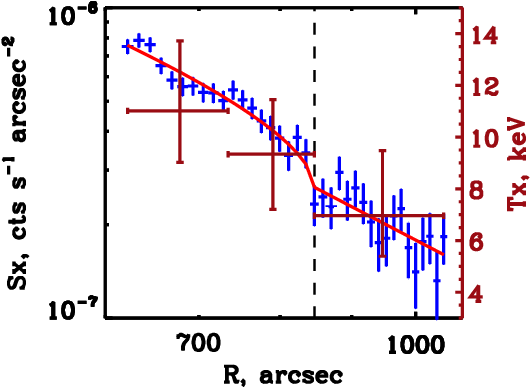}
\caption{
SBP and temperature profiles near the shock. The red solid line shows the best-fit broken power-law model. Red pluses are the temperature. The dashed line denotes the location of the SF, where a discontinuity in the density and temperature of the ICM is observed.
}
\label{fig:rp}
\end{center}
\end{figure}

The SBP in Figure~\ref{fig:rp} exhibits a subtle discontinuity, with a slight steepening of the profile at $R \approx 850''$. We fit the SBP with a broken power-law density model projected along the line of sight, following the method of \cite{Sarazin2016}. The best-fit model yields a density jump of $\rho_2/ \rho_1= 1.36 \pm 0.18$. Using the Rankine-Hugoniot jump condition \citep[e.g.,][]{Sarazin2016}, this corresponds to a shock Mach number of $\mathcal{M}_\rho = 1.24 \pm 0.13$.
Given the weakness of the shock, we further perform an independent check using the PROFFIT package \citep{Eckert2020}. The best-fit density jump from PROFFIT is \(1.38_{-0.12}^{+0.14}\), which yields a Mach number of \(\mathcal{M}_{\rho} = 1.26_{-0.08}^{+0.10}\). This result is fully consistent with our previous estimate, demonstrating that the shock detection is relatively robust.
We also extract spectra in two regions immediately inside and outside the edge (see Figure~\ref{fig:rp}) using the double-background subtraction method. The temperatures are $kT_2 = 9.3 \pm 2.1$ keV (post-shock) and $kT_1 = 7.4 \pm 2.0$ keV (pre-shock), which yields a Mach number of $\mathcal{M}_T = 1.34 \pm 0.60$, consistent with the density-derived value within the uncertainties. In the following discussion, we adopt $\mathcal{M}_i = 1.3$ as the shock Mach number in the ICM.
We note that this derived Mach number is likely a lower limit on the true shock Mach number, because shocks observed in X‑rays are often underestimated due to e.g., projection effects \citep[e.g.,][]{Akamatsu2017,Zhang2019}.
The detected shock front (SF) is marked by the cyan arc in Figure~\ref{fig:img}. It is located $\sim 1.3$ Mpc NW of the cluster center and close to the $R_{500}$, suggesting it is associated with a recent or ongoing merger event along the NW-SE axis, as supported by simulations discussed in the following Section~\ref{sec:scenario}.

\section{Discussion}
\label{sec:discussion}
\subsection{Merger scenario}\label{sec:scenario}

\begin{figure}
\begin{center}
\centering
\includegraphics[angle=0,width=0.45\textwidth]{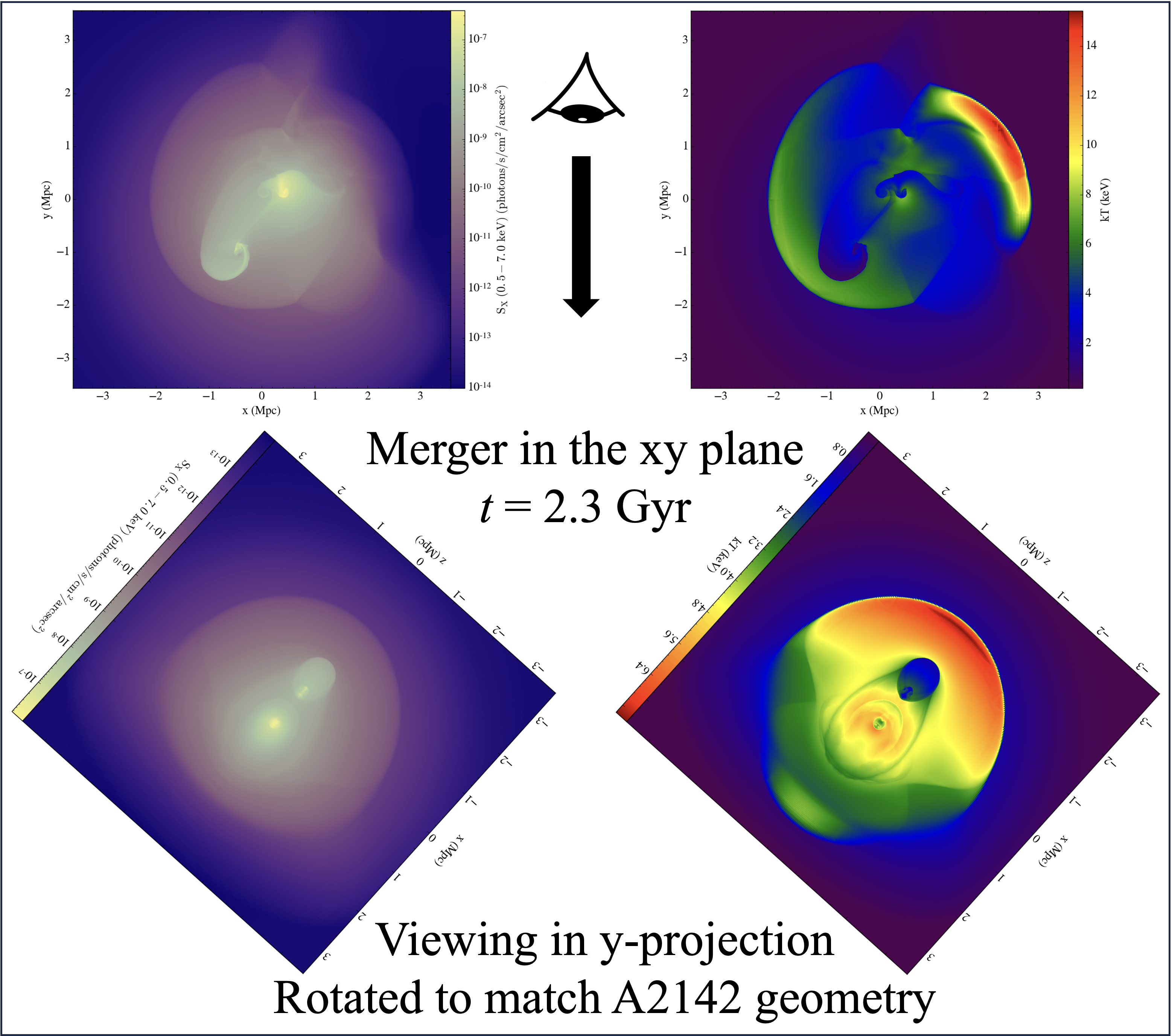}
\includegraphics[angle=0,width=0.45\textwidth]{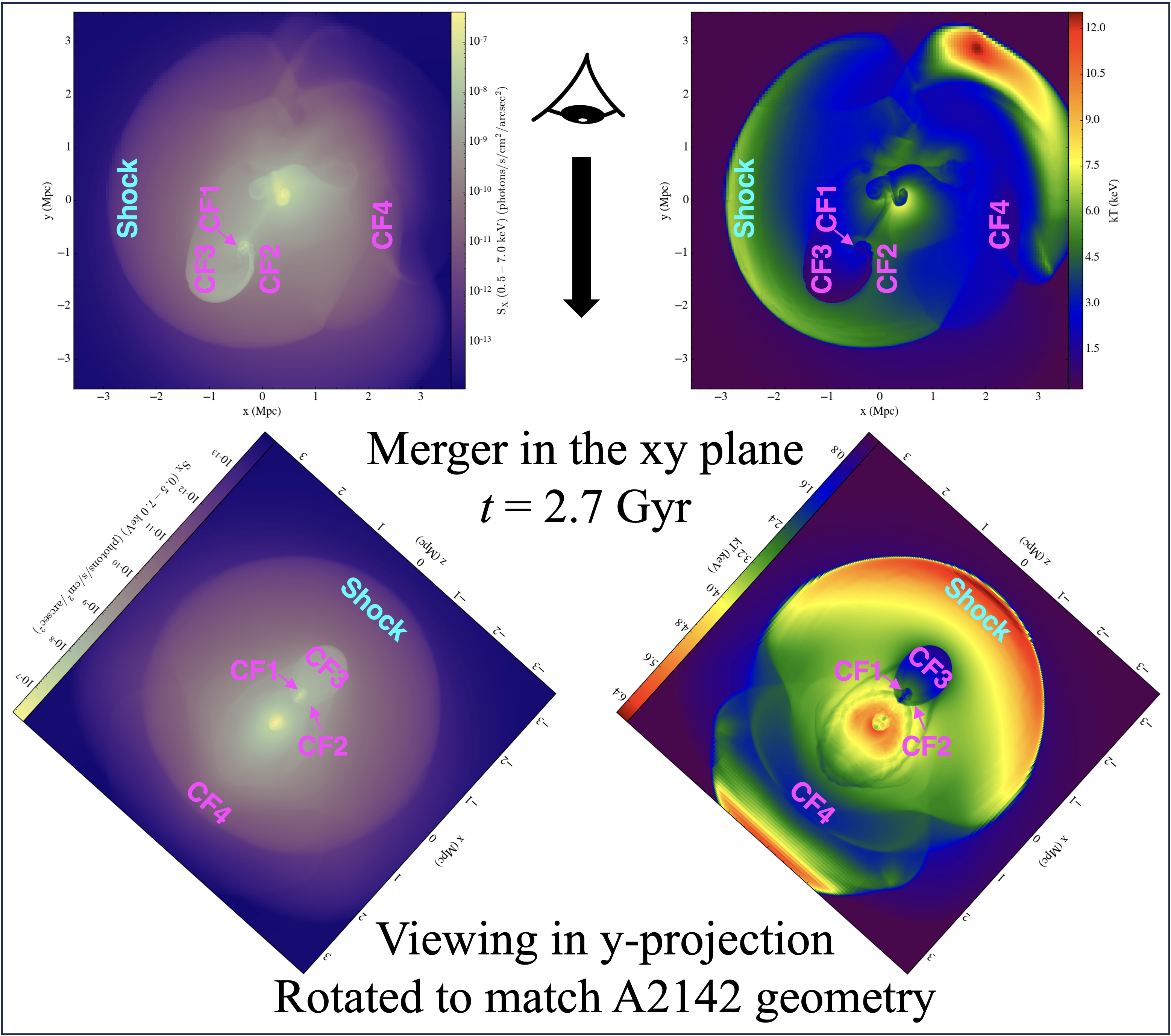}
\includegraphics[angle=0,width=0.45\textwidth]{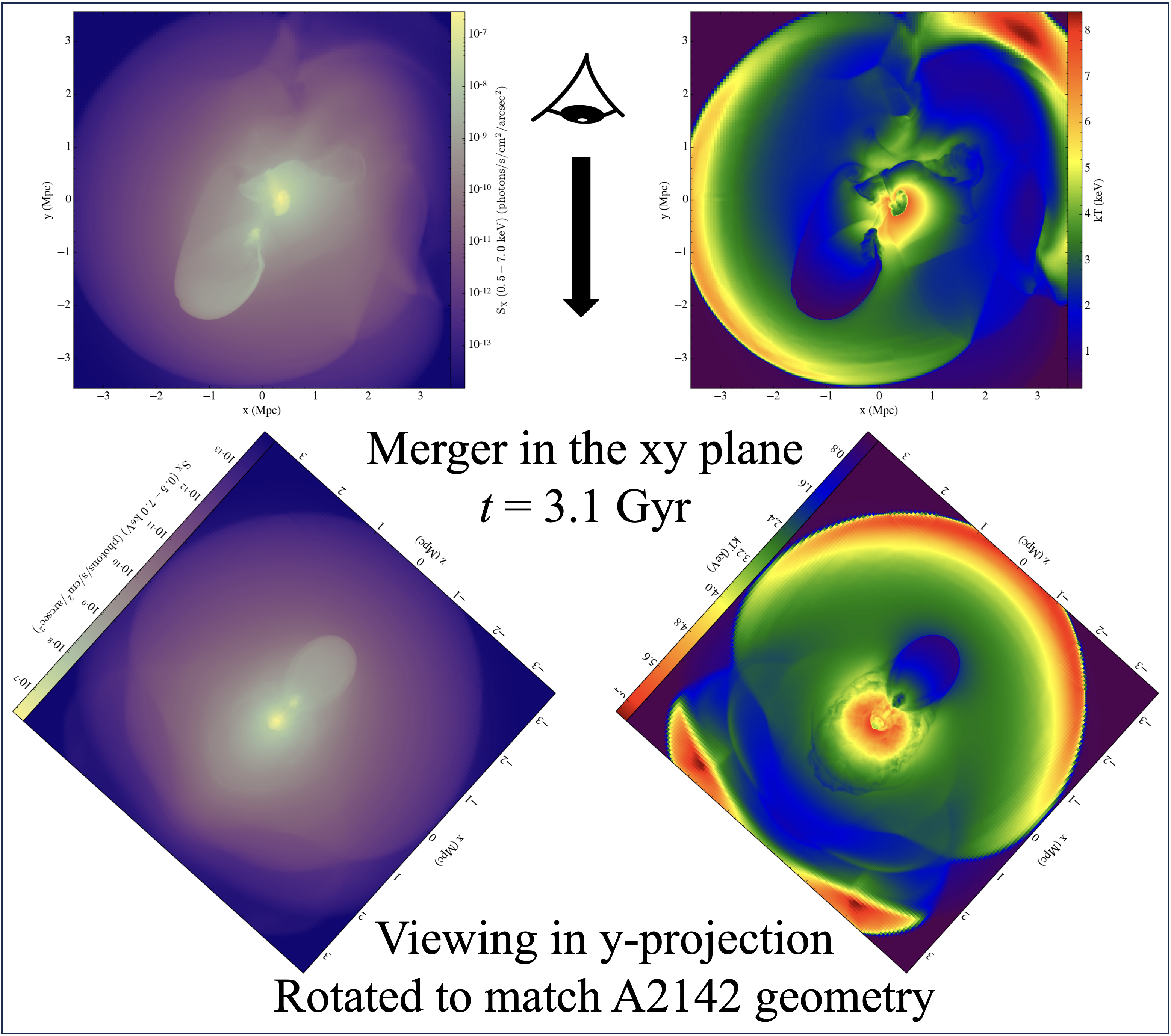}
\caption{
Simulations of galaxy cluster merger with a mass ratio $R=3$, initial impact parameter $b = 1000$ kpc, with different merger time $t$ from \cite{ZuHone2011}.
}
\label{fig:sim}
\end{center}
\end{figure}

The merger scenario of A2142 can be reproduced by numerical simulations of a galaxy cluster merger with a mass ratio $R = 3$, an initial impact parameter $b = 1000$~kpc, and a merger time $t = 2.7$~Gyr from \cite{ZuHone2011}, presented in Figure~\ref{fig:sim}. To illustrate the dynamic merger process and shock propagation, we also provide simulation snapshots at $t=2.3$~Gyr and 
3.1~Gyr in Figure~\ref{fig:sim}. Left panels are X-ray emissivity in 0.5-7 keV, and right panels are temperature maps. The relatively large impact parameter, exceeding the core radius of the primary cluster, results in an off-axis encounter that imparts significant angular momentum to the ICM of both subclusters. This is key to reproducing the complex morphology observed in A2142. 

As shown in Figure~\ref{fig:sim}, the merger occurs primarily in the $xy$-plane, but we are viewing it in the $y$-direction.
This viewing geometry is independently supported by the large line-of-sight velocity difference of $\sim 1600$~km~s$^{-1}$ between the two BCGs, and only one main mass peak of the cluster near BCG1 in the weak lensing map \citep{Okabe2008}. 
BCG2 (at $z = 0.09651$) is associated with the infalling subcluster and is receding from us, while BCG1 (at $z = 0.09081$), the cD galaxy of the primary cluster, resides near the cluster potential center and remains relatively undisturbed, consistent with the mild nature of the intermediate-mass-ratio merger.
The subcluster related to BCG2 is also identified as sub02 in the optical analysis of \cite{Liu2018}, and as S1 in \cite{Owers2011}, and both studies find that the velocity of the BCG2 subcluster relative to the main cluster is consistent with $\sim 1600$~km~s$^{-1}$. 
Meanwhile, the BCG2 is near BCG1 on the sky, and the weak lensing is sensitive to the total mass along the line of sight. If the subcluster associated with BCG2 is closely aligned with the main cluster (BCG1) along the line of sight as suggested by the simulation, weak lensing would only detect the main mass peak near BCG1.

The off-axis passage of the subcluster through the primary cluster does not destroy the cool core of subcluster, but instead triggers both the subcluster core sloshing and the large-scale sloshing of the ambient low-entropy gas. The sloshing motion of the subcluster core generates a series of concentric cold fronts at increasing 
radii, consistent with the three inner CFs (CF1, CF2, and CF3; \citealt{Markevitch2000,Wang2018}) observed in A2142 (Figure~\ref{fig:img}). While the large-scale sloshing propagates outward over time, and induces the outermost sloshing front CF4 reaches $\sim 1$~Mpc to the southeast, as one of the most distant cold front 
discovered by \cite{Rossetti2013}.
The corresponding positions of the four known CFs in Figure~\ref{fig:img} are also marked in the simulation image of Figure~\ref{fig:sim}.

Simultaneously, the infalling subcluster drives a bow shock ahead of it as it plunges through the ICM of main cluster, as shown in Figure~\ref{fig:sim}.
The Mach number derived from the BCG2 subcluster optical velocity difference of $\sim 1600$~km~s$^{-1}$ is $\mathcal{M}_o = \Delta v/a_i = 1600/1400 = 1.1$, 
where the ICM sound speed is $a_i = \sqrt{\frac{\gamma k T_1}{\mu m_p}} \approx 1400 \ {\rm km \ s^{-1}}$ for a pre-shock temperature $kT_1 = 7.4$ keV in A2142.
The optical Mach number $\mathcal{M}_o$ is generally consistent with the one from X-ray.
However, we note that this Mach number derived from the optical velocity is also a lower limit, because we ignore the velocity of BCG2 subcluster in the plane of the sky.
This shock propagates outward and is now located $\sim 1.3$~Mpc northwest of the cluster center. The simulation images in Figure~\ref{fig:sim} that viewed along the $y$-axis and rotated to match the geometry of A2142, show a clear correspondence between the simulated and observed features: the bullet core revealed in XMM-Newton residual image is the sloshing cool core of infalling subcluster, meanwhile the shock front and all four CFs are reproduced at positions consistent with the X-ray observations (see Figure~\ref{fig:img}). 
By comparing the shock positions at different merger times, the simulations show that the merger shock propagates outward in nearly all directions, it expands roughly spherically as it moves into the outskirts. This outward expansion is key to understanding its interaction with the radio galaxies. Because the shock expands outward in nearly all directions, the post-shock wind has a velocity component that pushes material radially away from the cluster center. When the shock passes the radio galaxies T1 and T2, the post-shock wind therefore drives their tails in the direction of the cluster outskirts. Even after accounting for projection, this model still predicts tails pointing towards the outskirts. The outward-directed wind naturally explains the observed tail orientation.

The temperature structure in the simulation (right panels of Figure~\ref{fig:sim}) also qualitatively matches the X-ray temperature measurements in A2142. The post-shock region exhibits elevated temperatures, while the sloshing core retains cooler gas. This is consistent with the higher temperature observed on the NW side of A2142 \citep[e.g.,][]{Henry1996,Owers2009,Tchernin2016} and the cooler gas associated with the central cold fronts.
The agreement between simulation and observations provides strong support for the intermediate-mass-ratio, off-axis merger scenario as the origin of the complex thermal and non-thermal structure of A2142.

\subsection{Shock traced by radio arcs}\label{sec:shockarc}

\begin{figure*}
\begin{center}
\centering
\includegraphics[angle=0,width=0.497\textwidth]{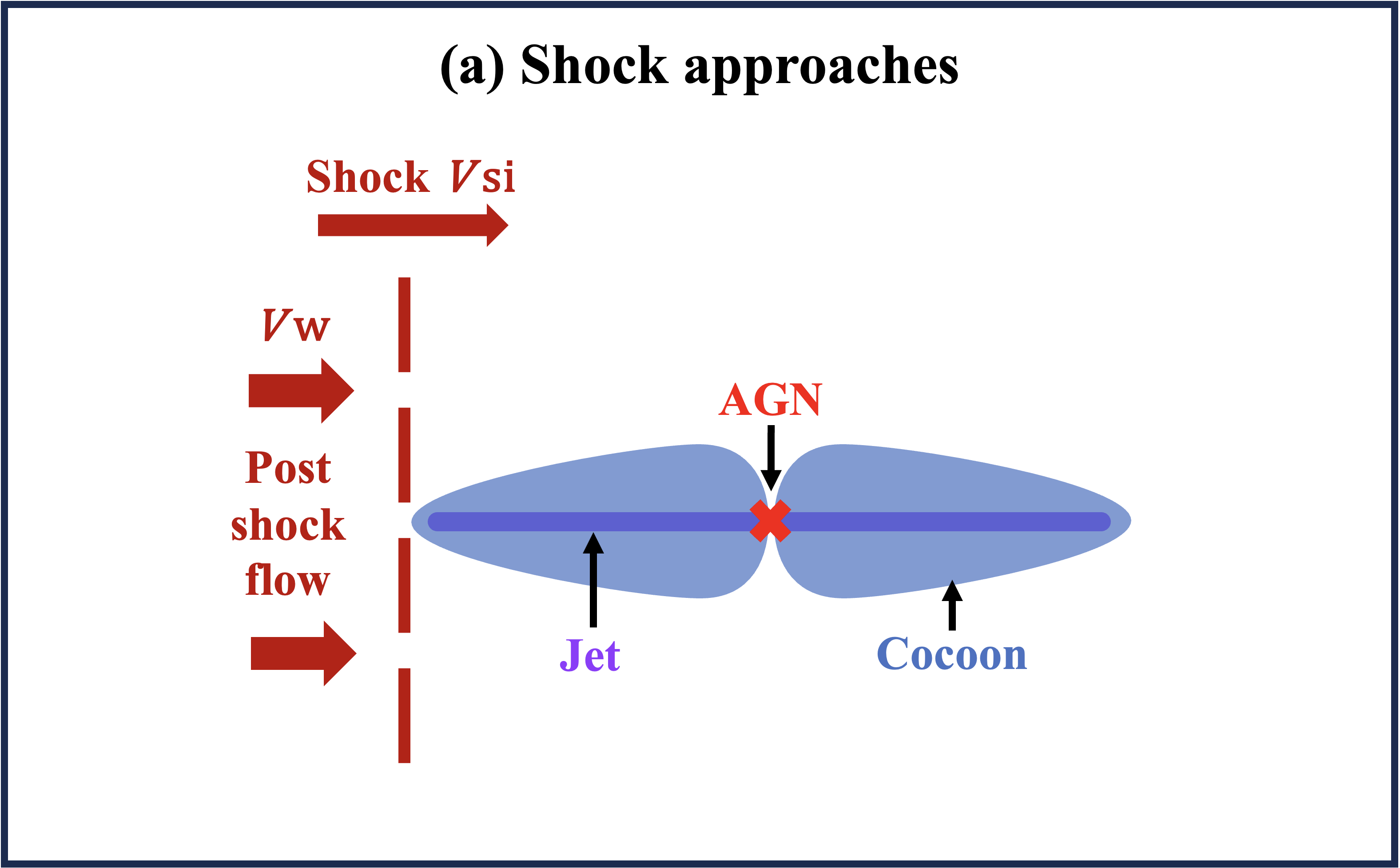}
\includegraphics[angle=0,width=0.497\textwidth]{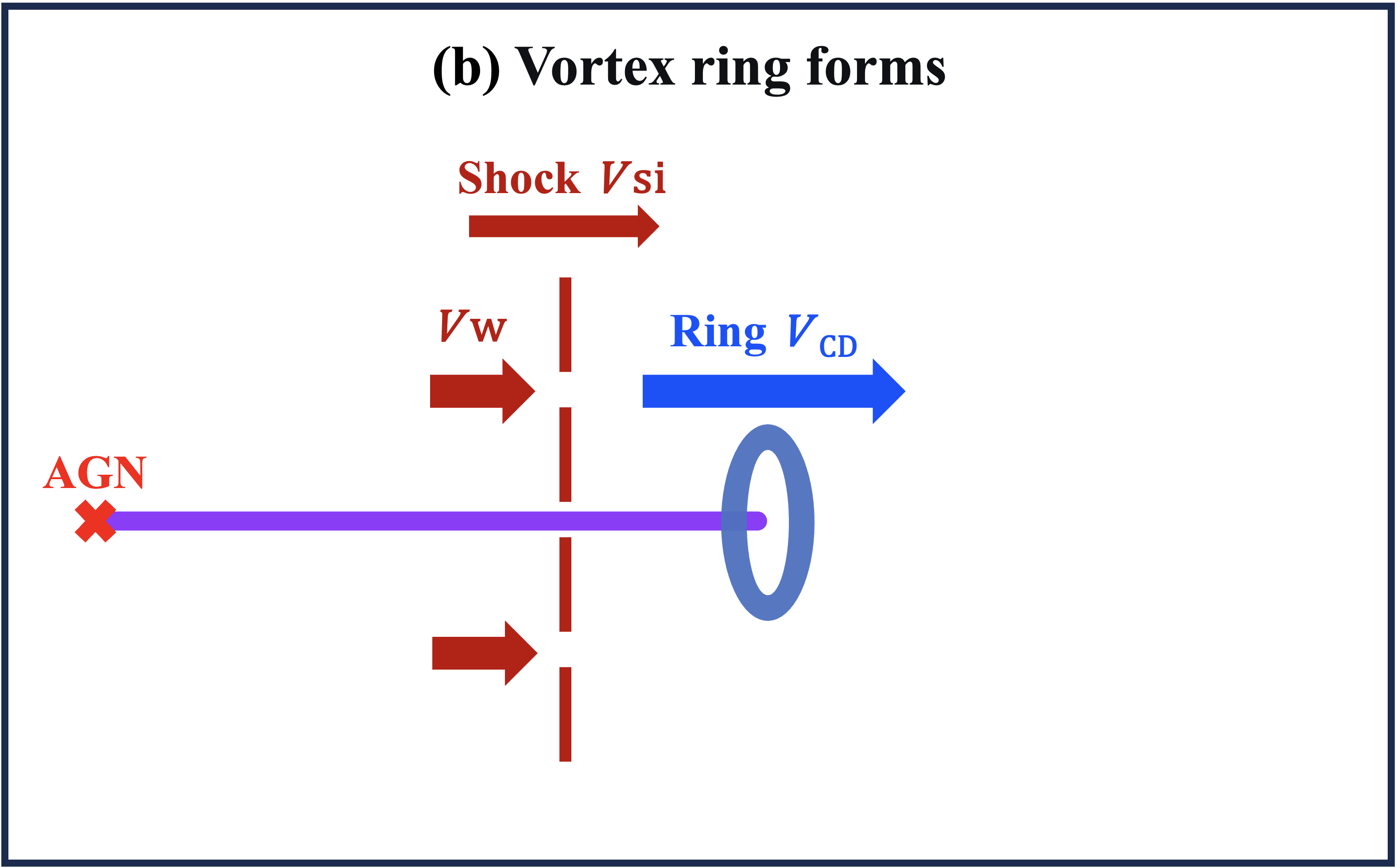}
\includegraphics[angle=0,width=0.497\textwidth]{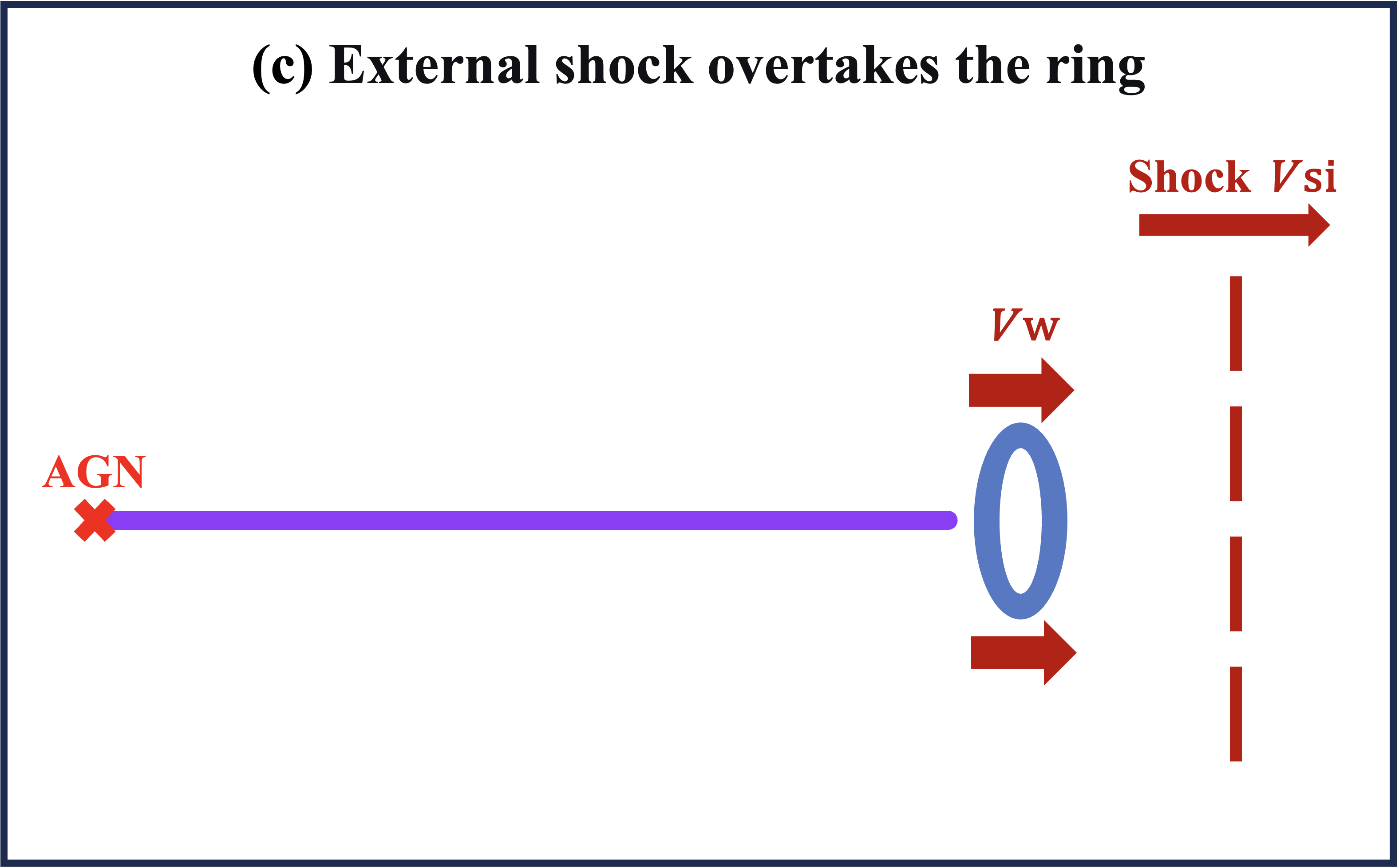}
\includegraphics[angle=0,width=0.497\textwidth]{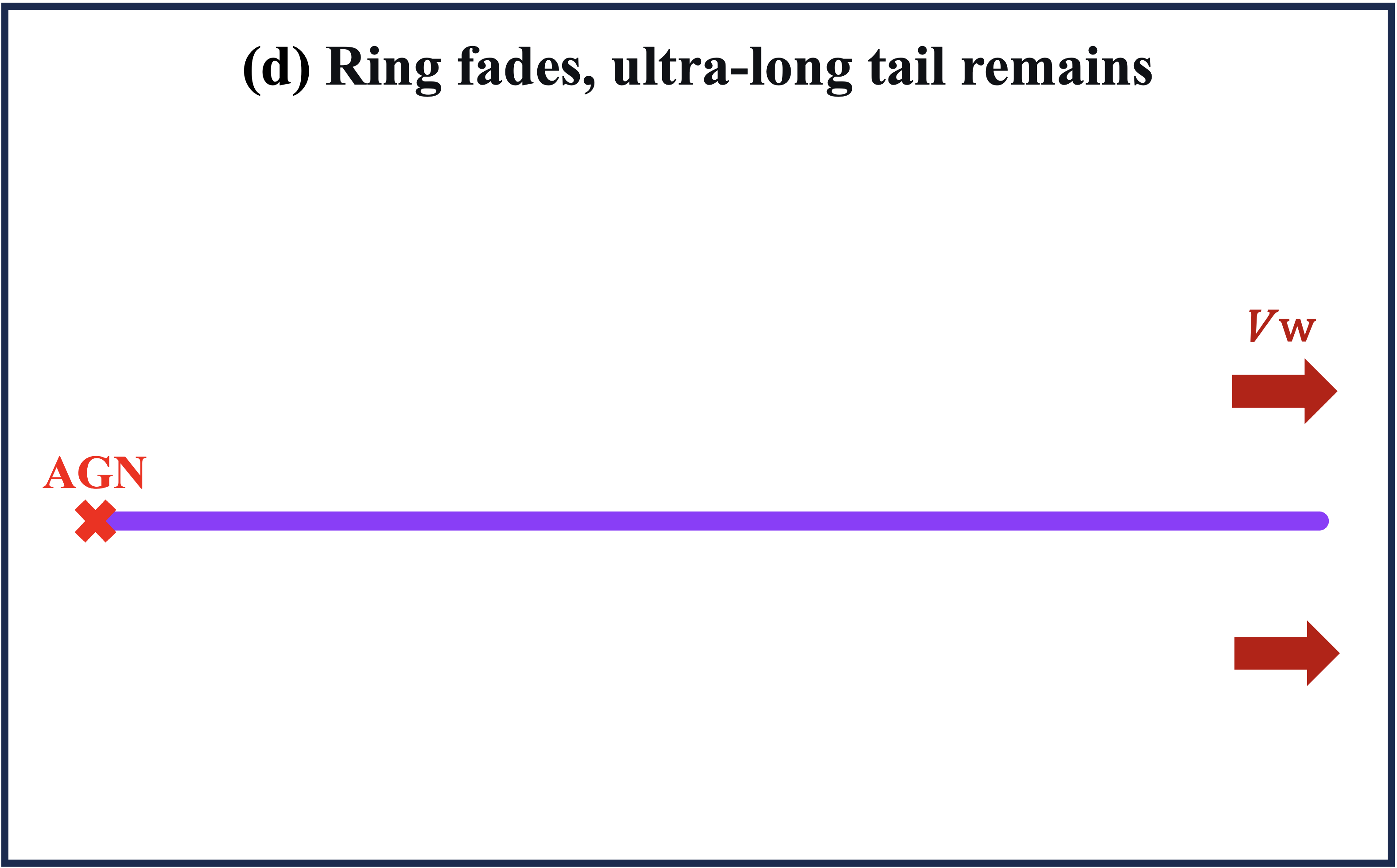}

\vspace{-0.1cm}
\caption{
Schematic illustration of the complete evolutionary
sequence of shock--radio galaxy interaction, based on the simulation of \cite{Nolting2019}.
{\bf (a)} The external shock (red dashed line) approaches a low-density cocoon (blue) surrounding the jet (purple solid line) of a radio AGN (red cross). The external shock velocity in ICM is $v_{\rm si}$ with a post-shock flow velocity of $v_w$.
{\bf (b)} Sufficiently strong post-shock winds can stop and even reverse the upwind jet, stripping it to a virtually naked state and leaving it without a surrounding cocoon. Due to the extreme density contrast ($\delta=\rho_{\rm cocoon}/\rho_{\rm ICM}\sim 0.01$) between the cocoon and ICM, the internal transmitted shock rushes through the cocoon at a much higher speed $v_{\rm sc} \gg v_{\rm si}$, and the related high speed post-shock flow in cocoon induces large shear flow and initiates contact discontinuity (CD) roll-up. The cocoon is crushed into a vortex ring within $\sim 100$ Myr. The ring, moving at a velocity comparable to the post-shock flow in the cocoon ($v_{\rm CD}$), will be decelerated by ram pressure toward the external post-shock wind speed $v_w$ as in panel (c).
While the relatively low speed external shock has not caught up with the ring, the ring leads the shock configuration as observed in A3627 and A2256, corresponding to this early stage.
{\bf (c)} The external shock, which propagates faster than the post-shock wind, eventually overtakes the ring. The current configuration in A2142, where the shock leads the ring, corresponds to this late stage.
{\bf (d)} As the vortex ring becomes disconnected from the AGN jet, no fresh cosmic‑ray electrons (CRe) are injected into it. The relativistic electrons in the ring cool radiatively, causing its radio emission to fade and its spectrum to steepen significantly. In contrast, the radio tail remains in contact with the AGN, new CRe can be continuously injected into the tail, maintaining a flatter spectrum. Eventually, at large distances from the AGN, even the tail may lose direct connection to the AGN, after which its spectrum also steepens. The shock‑stretched tail can persist as an ultra‑long ($>500$ kpc) fossil relic, observable at low radio frequencies as a steep‑spectrum structure.
}
\label{fig:cartoon}
\end{center}
\end{figure*}

The spatial alignment between the X-ray shock front and the arc-shaped radio filaments behind T1 and T2 strongly suggests a physical connection. We interpret these radio arcs as ``partial vortex rings" produced by the interaction of the merger shock with the low-density cocoon of the tailed radio galaxies, suggested by simulations \citep[e.g.,][]{Enblin2002,Jones2017,Nolting2019}. 
The complete evolutionary sequence of shock-cocoon interaction is illustrated in Figure~\ref{fig:cartoon}.
When the external merger shock encounters the low-density cocoon surrounding the radio jet, an internal transmitted shock is driven into the cavity. Because the cocoon density is orders of magnitude lower than the ambient ICM, the internal sound speed is extremely high. Consequently, the internal shock propagates through the cavity at a velocity $v_{\rm sc}$ far exceeding that of the external shock in the ICM ($v_{\rm si}$). This rapid internal penetration crushes the cocoon via intense shear along the cavity boundary, rolling it up into a toroidal vortex ring.
The cocoon crush timescale or vortex ring formation timescale $\tau_{\rm form}$ (typical $\tau_{\rm form} \sim 10^8$ yr; \citealt{Pfrommer2011}) is determined by the crossing time of the original cocoon-ICM contact discontinuity (CD) through the cocoon. 

Upon formation (Figure~\ref{fig:cartoon}b), the vortex ring initially acquires a speed comparable to that of the contact discontinuity, $v_{\rm CD}$, which is at the same speed as the post-shock flow in the cocoon, and intermediate between the internal shock speed and the external shock speed for $\mathcal{M}_i \gtrsim 2$ \citep{Pfrommer2011}.
However, vortex rings are known to self‑propagate faster than the surrounding medium due to their own induced velocity field (i.e., the circulation of the fluid; \citealt{Nolting2019b}). The magnitude of this self‑induced forward motion depends on the circulation, which is not well constrained in our case, but is likely a modest effect. Moreover, as a vortex ring propagates, it entrains material from the surrounding medium, which gradually slows its motion relative to the background post‑shock wind. The net effect is that the vortex ring may initially travel slightly faster than the post‑shock wind, but eventually decelerates.
In some circumstances, the cocoon may be crushed and stripped away from the propagating jet before the shock in the external medium propagates around it, as observed in A3627 (X-ray shock detected by \citealt{Ge2026}, and partial radio ring after tailed radio galaxy ESO~137-007 observed by e.g., \citealt{Koribalski2024}) and A2256 (X-ray shock detected by \citealt{Ge2020}, and source AG+AH as partial radio ring after tailed radio source C observed by e.g., \citealt{Osinga2024}).

Then, as shown in Figure~\ref{fig:cartoon}(c), the vortex ring  subsequently decelerates via ram pressure until it settles into the post-shock wind in ICM, drifting outward at velocity \citep[e.g.,][]{Jones2017}
\begin{equation}
v_w=\frac{3}{4}\frac{\mathcal{M}_i^2-1}{\mathcal{M}_i}a_i, 
\label{equ:vw}
\end{equation}
where $a_i$ is the sound speed of ICM. 
Because the post-shock wind speed $v_w$ is lower than the external shock speed $v_{\rm si}$, the external shock will eventually overtake the radio ring and move ahead of it, as observed in A2142. Based on the relative positions of the radio ring and the shock front, we can place a lower limit on the age of the radio ring.
With a shock Mach number $\mathcal{M}_i = 1.3$, the shock propagation velocity is
$v_{\rm si} = \mathcal{M}_i \, a_i \approx 1820 \ {\rm km \ s^{-1}}$, and 
the post-shock wind velocity is
$v_w \approx 560 \ {\rm km \ s^{-1}}$.
We note that the shock Mach number may evolve as it propagates through the stratified ICM; however, quantifying this evolution is not straightforward due to the varying density and temperature. Our estimate therefore adopts the current Mach number as a representative value, while acknowledging that the actual Mach number could have been different in the past.
The projected separation between the shock front and the vortex ring of T1 is $\Delta d_{\rm shock-ring} \approx 400$ kpc. The shock has been propagating ahead of the ring at a relative velocity of $v_{\rm si} - v_w \approx 1820 - 560 = 1260$ km s$^{-1}$. Ignoring projection effects, the time required for the shock to establish this lead after ring formation is:
\begin{equation}
t_{\rm dyn} \approx \frac{\Delta d_{\rm shock-ring}}{v_{\rm si} - v_w} \approx \frac{400 \ {\rm kpc}}{1260 \ {\rm km \ s^{-1}}} \approx 310 \ {\rm Myr}.
\end{equation}
This dynamical age ($t_{\rm dyn} \sim 310$ Myr) represents a lower limit on the time elapsed since the vortex ring was formed by the shock passage. 
Intriguingly, this value is in broad agreement with the radiative age of the electrons in the outer tail and arc regions of T1, estimated by \cite{Bruno2024} from spectral fitting to be $\sim 350$ Myr. 
This concordance between independent age estimates, one purely dynamical, the other from synchrotron aging, lends strong quantitative support to the shock--vortex ring model presented here.
The partial ring structure, combined with the steep radio spectrum ($\alpha \lesssim -1.5$; \citealt{Bruno2024}), indicates that we are observing aged relativistic electrons that have been adiabatically compressed and re-energized by the shock passage \citep[e.g.,][]{Enblin2002}.

\subsection{Shock elongates radio tail}\label{sec:shocktail}
In addition to the connection between shocks and radio arcs, a common feature of these head-tail radio galaxies with radio arcs is that their radio tails are very long ($>$500 kpc in A2142, A2256, and A3627).
Moreover, the velocities along the line of sight of T1 ($z = 0.09540$) and T2 ($z = 0.08953$) with respect to A2142 ($z = 0.08982$) are $\sim 1500$ km s$^{-1}$ and $\sim 80$ km s$^{-1}$, which indicates that their true tail lengths can be longer due to projection effects. 
We suggest that these ultra-long tails may also be attributed to the shocks.
Early work by \cite{Bliton1998} found that clusters with NAT radio galaxies exhibit a significantly higher level of X-ray substructure than radio-quiet clusters, indicating that NAT sources are preferentially located in dynamically complex, merging systems. 
In fact, cluster merger shocks can facilitate the formation of head-tail radio galaxies, because even a moderate-strength shock with $\mathcal{M}_i \sim 2$ can enhance ram pressure stripping by an order of magnitude \citep[e.g.,][]{Ge2026}. Depending on the shock--radio galaxy alignment, strong post-shock winds can either bend the jets into a NAT radio morphology or even reverse the upstream jet, producing a one-sided radio head-tail morphology \citep[e.g.,][]{Nolting2019,Nolting2019b}.

A fundamental constraint on the observable length of a radio tail is the radiative lifetime of the relativistic electrons. For a tail plasma advected from the host galaxy at a bulk velocity \(v_{\text{tail}}\), the maximum expected length set purely by synchrotron and inverse-Compton cooling is
\begin{equation}
    l_{\text{cool}} \approx v_{\text{tail}} \times t_{\text{cool}},
\end{equation}
where the cooling time \(t_{\text{cool}}\) at a given observing frequency \(\nu\) in a magnetic field \(B\) is estimated using Equation (3) from \citet{vanWeeren2019}:
\begin{equation}
    t_{\text{cool}} \approx 3.2 \times 10^{10} \frac{B^{1/2}}{B^2 + B^2_{\text{CMB}}} [(1+z)\nu]^{-1/2} \ \text{yr}.
\end{equation}
Here, \(B_{\text{CMB}} \approx 3.25 (1+z)^2\ \mu\text{G}\) is the equivalent magnetic field of the cosmic microwave background. For representative parameters of a tail at \(z = 0.1\) with a magnetic field \(B = 2\ \mu\text{G}\), observed at a frequency \(\nu = 100\ \text{MHz}\) (e.g., LOFAR) or \(\nu = 1\ \text{GHz}\) (e.g., VLA and MeerKAT), the cooling time $t_{\text{cool}}$ evaluates to 200 Myr for 100 MHz and 70 Myr for 1 GHz, respectively.
Due to efficient momentum exchange with the ambient ICM, the NAT plasma moves at a speed comparable to the surrounding wind. However, because the mixed tail plasma is less dense than the ICM, it receives a net velocity boost of roughly 10\% \citep{ONeill2019a}. Assuming a typical infall velocity of the host galaxy relative to the ICM of $\sim 900 {\rm\ km\ s^{-1}}$ (consistent with cluster velocity dispersion), the resulting tail bulk velocity is \(v_{\text{tail}} \approx 1000\ \text{km}\,\text{s}^{-1}\).
Then the corresponding cooling-limited length $l_{\text{cool}}$ is 200 kpc for 100 MHz and 70 kpc for 1 GHz.
The existence of tailed radio galaxies with lengths substantially exceeding this scale, reaching $>$500 kpc or more, presents a clear ``length problem''. Such extreme lengths demand a physical mechanism that either resets the radiative clock or continuously re-accelerates the emitting electrons.

Merger shocks can resolve this length problem through two interconnected processes that act in tandem: (1) the re-acceleration of aged electrons, and (2) the establishment of a high-speed post-shock wind that physically stretches the tail.
When a merger shock sweeps over an aging radio tail, the relativistic electrons are subject to adiabatic compression and diffusive shock acceleration, which re-energize the particle population \citep{Enblin2002, markevitch2007, vanWeeren2019}. This process allows the tail to shine brightly at large distances from the AGN core, far beyond the original \(l_{\text{cool}}\). Numerical magnetohydrodynamic simulations \citep[e.g.,][]{Nolting2019,Oneill2019} directly demonstrate this ``rejuvenation'' of NAT tails by passing shocks, showing that post-shock tails can be re-illuminated and extended to scales unattainable by passive aging.

The second, equally crucial, mechanism is the kinematic stretching of the tail. As discussed in Section~\ref{sec:shockarc} (Equation~\ref{equ:vw}), the passage of a merger shock establishes a post-shock wind in the ICM that streams at a velocity \(v_w\).
For the shock in A2142 (\(\mathcal{M} \approx 1.3\), \(a_i \approx 1400\) km/s), this yields \(v_w \approx 560\) km/s. For a moderate shock with \(\mathcal{M} \sim 2\), \(v_w\) can exceed 1500 km/s. The old, re-accelerated plasma is entrained in this wind and advected downstream at high speed. Consequently, the total achievable tail length becomes:
\begin{equation}
    l_{\text{tail}} \sim (v_{\rm tail}+v_w) \times t'_{\text{cool}},
\end{equation}
where \(t'_{\text{cool}}\) is the renewed cooling time of the re-accelerated electrons. Since both $(v_{\rm tail} + v_w)$ and \(t'_{\text{cool}}\) are substantially larger than their values in the unshocked medium, the tail can be inflated to multiple times its original cooling-limited length.
This dual role of the shock, as a particle re-accelerator and a high-speed conveyor, provides a unified explanation for the Mpc-scale tails observed in merging clusters, as depicted in Figure~\ref{fig:cartoon}(d).

The physical arguments presented above lead to a key methodological implication: ultra-long radio tails (\(l_{\rm tail} \gg l_{\text{cool}}\)) can serve as an independent, complementary tracer of merger shocks in galaxy clusters. 
Thus, radio galaxies can act as ``weather vanes'' of the stormy weather driven by shocks and turbulence in galaxy clusters \citep[e.g.,][]{Burns1998}.

\section{Conclusions}

We have presented the discovery of a merger shock on the northwest side of the merging galaxy cluster A2142 using deep XMM-Newton observations. Our main findings are as follows:

\begin{enumerate}
    \item Merger shock detection: We detect a subtle surface brightness discontinuity and a corresponding temperature jump on the NW side of A2142 at a projected distance of approximately 1.3~Mpc from the cluster center. The density jump yields a shock Mach number of $\mathcal{M}_{\rho} = 1.24 \pm 0.13$, and the temperature jump yields $\mathcal{M}_{T} = 1.34 \pm 0.60$. This merger shock is consistent with earlier indications of shock heating and SZ pressure enhancement in the NW region. 
    The observed shock front and four known cold fronts are well reproduced by numerical simulations \citep{ZuHone2011} of an off-axis merger with a mass ratio $R = 3$ and a large impact parameter $b = 1000$~kpc, which imparts significant angular momentum to the ICM of both subclusters, triggering the sloshing motions that generate the multiple CFs and driving the outward-propagating shock.

    \item Shock--vortex ring connection: The X-ray shock front is spatially coincident with arc-shaped radio filaments extending beyond the tails of the head-tail radio galaxies T1 and T2. We interpret these radio arcs as partial vortex rings, like ``smoke rings'', produced by the interaction of the merger shock with the low-density cocoons of the radio galaxies. The steep radio spectrum and the relative geometry of the shock and rings support a scenario in which aged relativistic electrons are adiabatically compressed and re-energized by the shock passage.

    \item Shock--ring relative position as an evolutionary clock: The relative ordering of the external shock front and the vortex ring provides a diagnostic of the merger stage. In A2256 \citep[e.g.,][]{Ge2020,Osinga2024} and A3627 \citep[e.g.,][]{Koribalski2024,Ge2026}, the vortex ring lies ahead of the external shock front (ring-ahead configuration), indicating a recent encounter in which the shock has not yet had sufficient time to overtake the ring. In contrast, in A2142, the shock front leads the vortex ring (shock-ahead configuration), indicating a more evolved state where the external shock has already propagated ahead of the decelerated ring. From the projected separation between the shock and the ring in A2142, we derive a lower limit on the dynamical age of the ring of approximately $310$~Myr, in broad agreement with the synchrotron radiative age of the electrons in the outer tail and arc regions of T1.

    \item Shock elongation of radio tails and a new shock diagnostic: We demonstrate that merger shocks can significantly elongate radio tails through two interconnected mechanisms: the re-acceleration of aged relativistic electrons and the kinematic stretching of the tail plasma by the post-shock wind. The combination of these effects can inflate radio tails to lengths far exceeding the nominal synchrotron cooling length, resolving the ``length problem'' for the $>$500~kpc tails observed in A2142, A2256, and A3627. This implies that ultra-long radio tails can act as independent, complementary tracers of merger shocks in galaxy clusters, even in the absence of deep X-ray data. 
\end{enumerate}

In summary, the discovery of the merger shock in A2142, together with the analogous cases of A2256 and A3627, establishes two independent and complementary radio-based diagnostics of merger shocks in galaxy clusters. First, arc-shaped radio filaments, the partially deformed vortex rings of stripped and rolled jet cocoons, carry the characteristic signatures of shock--cocoon interactions: steep spectra, curved morphologies, and a spatial offset from the host galaxy's tail that encodes the evolutionary stage of the shock encounter. Second, ultra-long radio tails ($l_{\mathrm{tail}} \gg l_{\mathrm{cool}}$), inflated by the combined action of shock-induced particle re-acceleration and kinematic stretching by the post-shock wind, serve as independent tracers of past shock passages. Crucially, neither diagnostic requires deep X-ray data for shock identification: the radio morphology, spectral properties, and tail length alone provide sufficient information to infer the presence of a shock, its approximate strength and direction, and its evolutionary stage. In this sense, tailed radio galaxies function as ``cluster weather vanes'', and both their vortex ring remnants and their elongated tails constitute a fossil record of ICM storms. Future high-resolution, low-frequency radio surveys with instruments such as LOFAR and the Square Kilometre Array will uncover many more such systems. A systematic census of radio arcs and ultra-long tails will provide a novel, radio-selected probe of the frequency, strength, and geometry of merger shocks across the local Universe, independent of X-ray selection biases.

\begin{acknowledgments}
This research has made use of data and/or software provided by the High Energy Astrophysics Science Archive Research Center (HEASARC).
CG acknowledges support from the National Natural Science Foundation of China (No. 12373007, 12422302).
\end{acknowledgments}

\bibliography{master}{}
\bibliographystyle{aasjournalv7}

\end{document}